\documentclass{aa}
\usepackage{graphicx}
\usepackage[varg]{txfonts}
\begin{document}
   \title{Non-thermal radio emission from O-type stars}
   \subtitle{I. HD~168112}

   \author{R. Blomme \inst{1} \and
           S. Van Loo \inst{1} \and M. De Becker \inst{2} \and
           G. Rauw \inst{2}\thanks{Research Associate FNRS (Belgium)}
           \and M.C. Runacres \inst{1}
           \and D. Y. A. Setia Gunawan \inst{3}
           \and J.M. Chapman \inst{3}
          }

   \offprints{R. Blomme, \\ \email{Ronny.Blomme@oma.be}}

   \institute{Royal Observatory of Belgium,
              Ringlaan 3, B-1180 Brussel, Belgium
            \and
              Institut d'Astrophysique, Universit\'e de Li\`ege,
              All\'ee du 6 Ao\^ut, B\^at B5c,
              B-4000 Li\`ege (Sart-Tilman), Belgium
            \and
              Australia Telescope National Facility,
              PO Box 76, Epping, NSW 2121, Australia
             }

   \date{Received date; accepted date}

   \abstract{
We present a radio lightcurve of the O5.5 III(f$^+$) star HD~168112,
based on archive data from the Very Large Array (VLA) and the Australia 
Telescope Compact Array (ATCA).
The fluxes show considerable variability and a negative spectral index,
thereby confirming that HD~168112 is a non-thermal radio emitter.
The non-thermal radio emission is 
believed to be due to synchrotron radiation
from relativistic electrons that have been Fermi accelerated in
shocks. For HD~168112, it is not known whether these shocks are due to
a wind-wind collision in a binary system or to
the intrinsic instability of the stellar wind driving mechanism.
Assuming HD~168112 to be a single star, 
our synchrotron model shows
that the velocity jump of the shocks should be very high, or there
should be a very large number of shocks in the wind. Neither
of these is compatible with time-dependent hydrodynamical calculations
of O star winds.
If, on the other hand, we assume that HD~168112 is a binary, the high
velocity jump is easily explained by ascribing it to the wind-wind
collision. By further assuming the star to be an eccentric binary,
we can explain the observed radio variability by the colliding-wind
region moving in and out of the region where 
free-free absorption is important. The radio data presented here
show that the binary has a period of between one and two years.
By combining the radio data with X-ray data, we find that 
the most likely period is $\sim$~1.4~yr.
   \keywords{stars: individual: HD 168112 --
             stars: early-type -- 
             stars: mass-loss -- 
             stars: winds, outflows -- 
             radio continuum: stars -- 
             radiation mechanisms: non-thermal
               }
   }

   \maketitle

\section{Introduction}
\label{section introduction}

Radio emission from most hot stars is due to the thermal free-free emission
by material in their stellar winds. A significant fraction of these stars,
however, also show evidence of non-thermal emission. 
The radio fluxes of non-thermal sources are
characterized by a zero or negative spectral 
index\footnote{The radio spectral index $\alpha$ is
      defined by $F_\nu \propto \nu^\alpha \propto \lambda^{-\alpha}$.
      For thermal emission $\alpha \approx +0.6$}, 
significant variability and a high brightness temperature
(Bieging et al.~\cite{Bieging+al89}).
White (\cite{White85}) attributed the non-thermal radiation to
relativistic electrons that spiral in a magnetic field and thereby
emit synchrotron radiation.
These electrons are accelerated
by the first-order Fermi mechanism (Bell~\cite{Bell78})
in the shocks present in the stellar wind. 
In a binary, the relativistic electrons are accelerated
in both shocks formed by the colliding winds (Eichler \&
Usov~\cite{Eichler+Usov93}).
In a single star, the shocks are due to the instability of the radiative
driving mechanism (e.g. Owocki et al.~\cite{Owocki+al88}). 

For Wolf-Rayet stars, Dougherty \& Williams (\cite{Dougherty+Williams00}) 
showed that non-thermal emission is strongly correlated with binarity:
7 out of the 9 non-thermal WR emitters are identified as binaries. 
Not all WR binaries are non-thermal emitters: if the period is
sufficiently short ($P \la 1$ yr), the synchrotron emission
is completely absorbed by the high free-free opacity
in these stellar winds.

For O stars, the link between non-thermal emission and binarity is less 
clear. While colliding winds are an explanation in some cases,
there remain a number of apparently single O stars that are
non-thermal emitters. Relativistic electrons in these stars
would then be accelerated in wind-embedded shocks. 
Time-dependent hydrodynamical
calculations confirm that the instability in the radiative driving
mechanism results in a significant number of shocks
(e.g. Runacres \& Owocki~\cite{Runacres+Owocki05}, and references therein).
Models for non-thermal emission of O-type stars, based on this mechanism,
have been developed by Chen \&
White~(\cite{Chen+White94}) and
Van Loo et al.~(\cite{VanLoo+al04}, \cite{VanLoo+al05}).

The purpose of the present paper is to study the non-thermal radio emission
of the apparently single star \object{HD~168112}. We do this by collecting
all available archive data. We explore both the single-star and binary
scenario, to see which one best explains the characteristics of the
non-thermal emission. The main characteristics we study are the 
radio flux levels,
spectral shape and time-scale of the variability.

\object{HD~168112}
belongs to the \object{Ser OB 2} association (Cappa et al.~\cite{Cappa+al02})
and lies inside \object{NGC 6604}, located at the core of the H II
region S54 (Georgelin et al.~\cite{Georgelin+al73}). 
The stellar parameters and wind parameters 
used in this paper are listed in
Table~\ref{table stellar parameters}.
De Becker et al.~(\cite{De Becker+al04}) classified \object{HD~168112}
as O5.5 III(f$^+$). They also showed that optical spectra 
do not reveal radial velocity variations: thus, \object{HD~168112} is 
not a binary, or it has a low inclination orbit or a long period.

Abbott et al.~(\cite{Abbott+al85}) were the first to identify
\object{HD~168112} as a non-thermal emitter.
Radio observations indicate
a zero or negative spectral index as well as
considerable variability.
Moreover, even at minimum, its
6~cm radio flux is a factor of 10 higher than that expected from
free-free radiation of the mass loss rate as determined from H$\alpha$,
which is a further indicator of non-thermal emission
(Bieging et al.~\cite{Bieging+al89},
De Becker et al.~\cite{De Becker+al04}).

De Becker et al.~presented two radio observations separated by 6 months,
showing a factor of 5--7 increase. 
Quasi-simultaneous XMM-Newton
observations show anti-correlated behaviour: there is a decrease
of $\sim 30$~\% in the X-ray flux.
De Becker et al.~interpret 
such behaviour as pointing to a binary scenario, where a significant fraction
of the X-ray emission is produced by the hot plasma heated by
the wind-wind interaction.
The present paper discusses a much larger number of radio observations.
These will allow us to quantify the non-thermal properties of 
\object{HD~168112}
and thus to better evaluate the merits of the single-star versus
binary scenario.

\begin{table}
\caption{The parameters of \object{HD~168112} from the SIMBAD catalogue,
Leitherer~(\cite{Leitherer88}), Bieging et al.~(\cite{Bieging+al89})
and Conti \& Ebbets~(\cite{Conti+Ebbets77}).}
\label{table stellar parameters}
\begin{center}
\begin{tabular}{llllll}
\hline
\hline
RA (J2000) & $18^{\rm h}18^{\rm m}40{\fs}868$ \\
Dec (J2000) & $-12{\degr}06{\arcmin}23{\farcs}37$ \\
$T_{\rm eff}$ & 46\,500 K \\
log L$_{\rm bol}$/L$_{\sun}$ & 6.06 \\
R$_*$ & 16 R$_{\sun}$ \\
M$_*$ & 70 M$_{\sun}$ \\
$v_{\infty}$ & 3250 km s$^{-1}$ \\
$\dot{M}$ & $2.5 \times 10^{-6}$ M$_{\sun}$ yr$^{-1}$ \\
distance & 2 kpc \\
$v_{\rm rot} \sin i$ & 90 km s$^{-1}$ \\
\hline
\end{tabular}
\end{center}
\end{table}

The remainder of this paper is structured as follows.
In Sect.~\ref{section data} we present the data. 
In Sect.~\ref{section interpretation} we 
explore both the single-star and binary scenario and
in Sect.~\ref{section conclusions} we present our conclusions.

\section{Data}
\label{section data}

Tables~\ref{table UCX data} and \ref{table LP data}
list the radio observations centred on, or near to, \object{HD~168112}.
These data were collected from the archives of the NRAO Very Large
Array\footnote{The National Radio Astronomy
             Observatory is a facility of the National Science Foundation
             operated under cooperative agreement by Associated Universities,
             Inc.}
(VLA) and the Australia Telescope Compact 
Array\footnote{The Australia Telescope Compact Array is part of the
          Australia Telescope which is funded by the Commonwealth of
          Australia for operation as a National Facility managed by 
          CSIRO.}
(ATCA).
Many of these observations have not been published previously.
To avoid introducing systematic shifts between different data sets,
we decided to re-reduce all observations in a consistent way.
Details of the reduction and the measured fluxes are given
in Appendix~\ref{section data reduction}.

\subsection{Long-term variability}
\label{section long-term variability}

\begin{figure*}
\resizebox{\hsize}{!}{\includegraphics[bb=28 50 509 537]{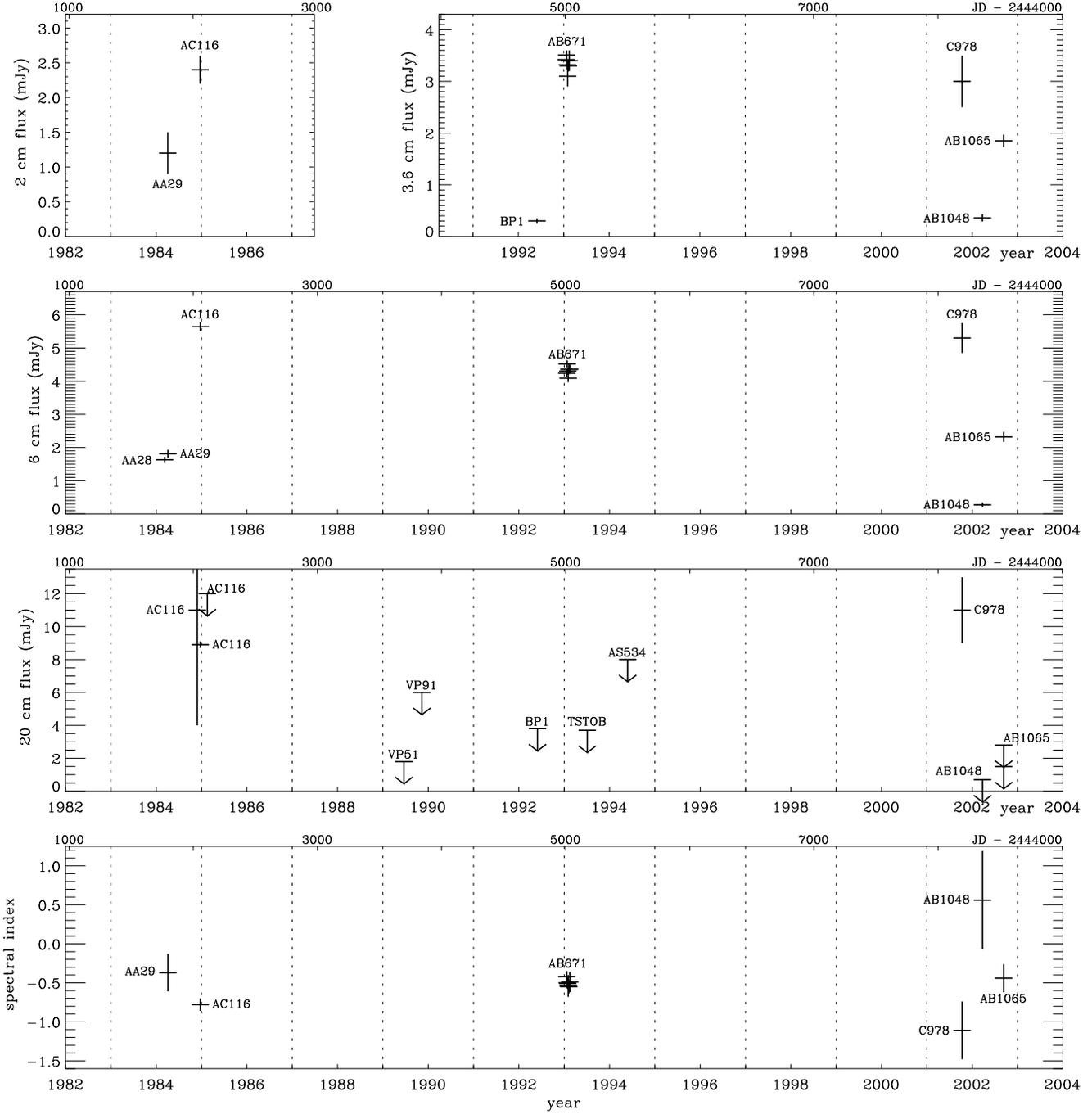}}
\caption{The 2, 3.6, 6 and 20~cm fluxes of the VLA and ATCA archive data,
plotted as a function of time. The bottom panel shows the spectral
index between 3.6 and 6~cm (or 2 and 6~cm if there is no 3.6 cm
observation). Observations are
labelled by their programme name. 
Only those 20~cm upper limits below 13~mJy have been included.}
\label{figure fluxes}
\end{figure*}

The VLA and ATCA fluxes at 2, 3.6, 6 and 20~cm are plotted as a function of 
time in Fig.~\ref{figure fluxes}.
Not all 20~cm upper limits were included in this figure, only those which
are lower than 13 mJy. 
The single 13~cm detection is not plotted.
The 90~cm observations are not shown as they only provide upper limits.
Note that there may be slight differences in wavelength for observations
we classed in the same wavelength band. 
The exact frequencies for each observation
are listed in Table~\ref{table calibrators}.

The detections at 2, 3.6 and 6~cm in Fig.~\ref{figure fluxes} clearly
show the variability that is an identifying characteristic of non-thermal
emission. Observations differing by more than 0.4 yr show significantly
different fluxes. At certain times, 
\object{HD~168112} is in a low state (around year 1992.4 and 2002.2),
while at other times it is in a high state (1985.0, 1993.1 and 2001.8). 
The flux differs by a factor of $\sim 20$ between these states.
At 1984.2 and 2002.9, the fluxes are intermediate between the two states. 
Although few detections have been made at 20~cm,
there are some defining a high
state (1985.0, 2001.8) and some of the upper limits are sufficiently
sharp to confirm the presence of a low state (2002.2).

The flux variations are not due to the fact that we are resolving the
wind of the star in the highest-resolution configurations. A 
high-resolution observation typically has a beam of 0{\farcs}5~diameter,
corresponding to $\sim$~13\,000 R$_*$ at 2 kpc distance. This value is much
higher than the expected size of a synchrotron emitting region, either in
the single-star or binary scenario.
We also note that the 20~cm images contain
another source (\object{GPS G018.565+01.756}, at $\sim 11\arcmin$
from \object{HD~168112}):
the flux of this strong source ($\sim 50$~mJy) varies by only 20~\%
between the \object{HD~168112} high (1985.0) and low (2002.2) state. 
We therefore consider that an instrumental effect is unlikely to be the
explanation for the observed variability.

Figure~\ref{figure fluxes} also shows that the spectral
index between 3.6 (or 2) and 6~cm is negative in most cases.
This is a further indication
that the emission is non-thermal. The only exception is at 2002.2,
where there is a large error bar that includes the thermal value 
($\alpha=+0.6$).
This does not mean, however, that we are observing the underlying free-free
emission of the wind. As De Becker et al.~(\cite{De Becker+al04})
already pointed out, the expected thermal flux is only 0.03~mJy at 6~cm,
which is a factor of 10 lower than the observed flux.
We further note that the most extreme values of the spectral index occur
when the most extreme values of the flux are reached. When the
flux is high, we find a very negative spectral index
($\alpha = -1.1 \pm 0.4$ at 2001.1 and
$\alpha = -0.78 \pm 0.08$ at 1985.0).
When the fluxes are low (2002.2), we find $\alpha = +0.6 \pm 0.6$.

The spectral index between 6 and 20~cm provides little additional
information and is therefore not shown in Fig.~\ref{figure fluxes}.
It is flatter than the 2--6~cm one at 1985.0.
Around 2002.9, the 6--20~cm index is positive,
indicating a turnover in the spectrum. Other observations
do not allow a significant determination of the 6--20~cm
spectral index.

\subsection{Short-term variability}
\label{section short-term variability}

\begin{figure}
\resizebox{\hsize}{!}{\includegraphics[bb=28 56 268 339]{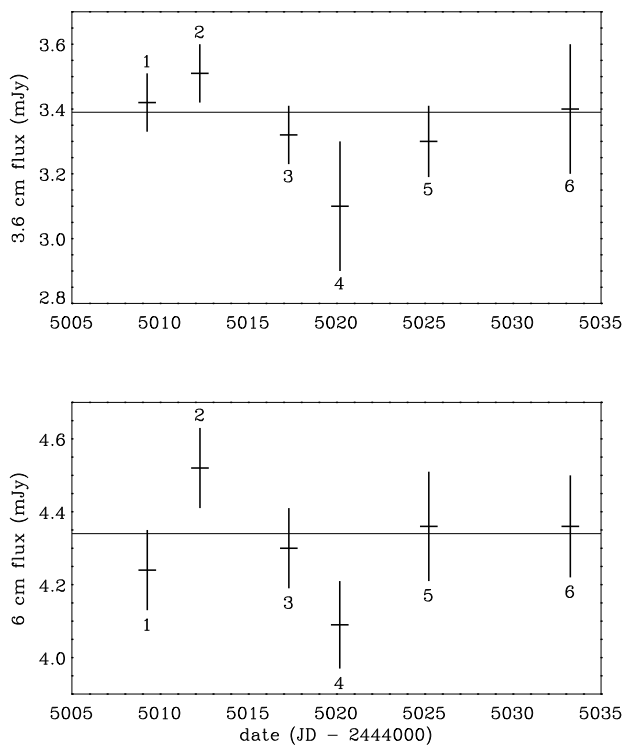}}
\caption{The 3.6 and 6~cm fluxes of the AB671 observations.
The solid line is the flux measured on the combined dataset.}
\label{figure AB671 fluxes}
\end{figure}

The shortest time-scales are covered by the observations
with programme name AB671 (around 1993.1). 
These data consist of 6 observations
at both 3.6 and 6~cm, spread roughly uniformly over $\sim$~0.1 yr
(1993 January 21 to February 14).
Figure~\ref{figure AB671 fluxes} shows that
observations No.~2 and 4 deviate from the average by 5--10~\%.
While this may suggest variability, it will be shown
that this variability is spurious.

A typical test to detect an instrumental cause for apparent variability
is to measure another source on the same image and see
whether any flux changes are correlated to the object
changes.
Unfortunately, only one additional source
is visible on the 3.6 and 6~cm images 
(at RA(J2000)=$18^{\rm h}18^{\rm m}40^{\rm s}$,
Dec(J2000)=$-12${\degr}07\arcmin) and it is too weak ($\sim 0.5$ mJy)
to see if its flux changes are correlated with those of
\object{HD~168112}.

A possible cause of apparent variability is the presence of large 
atmospheric phase changes that are not sufficiently well corrected for.
In the reduction of the AB671 dataset, self-calibration was applied
to follow the phase changes on a shorter time-scale than the phase calibrators 
allow (Appendix~\ref{section AB671 self-calibration}).
Because observation No.~4 had the lowest flux in the standard reduction
(i.e. without self-calibration), we suspect that in this specific case 
significant phase changes occurred on shorter time-scales than could be
corrected for with self-calibration.
The fact that the flux for No. 4 is low is therefore
not significant. 

Observation No. 2 deviates from the solid line
in Fig.~\ref{figure AB671 fluxes} by $1.33$~$\sigma$ at 3.6~cm
and $1.64$~$\sigma$ at 6~cm. A value of $1.64$~$\sigma$ corresponds to
a 10 \% (two-sided) probability of a Gaussian distribution. As we have
6 observations, it is not significant that one of them deviates
by $1.64$~$\sigma$.
We therefore conclude that we find no evidence for significant variability in
the AB671 dataset.
This allows us to combine all AB671 data into a single dataset
and apply our reduction technique to it. The result for
both wavelengths is given in
Table~\ref{table UCX data} and shown as a solid line in
Fig.~\ref{figure AB671 fluxes}.

Some of the other data allow us to look for variability on the 0.1--0.2~yr
time-scale. The two 6~cm observations AA28 and AA29 (at 1984.2)
are $\sim$ 0.1~yr apart and the flux is constant, within the
error bars.
There are three AC116 20~cm observations (around 1985), which cover
a $\sim$ 0.2~yr time-scale. Unfortunately,
one of the observations only gives a (high) upper limit and another
has a large error bar. No significant information about variability
can be derived from these data.

We do find clear variability on the time-scale of
0.45~yr between the C978 (2001.8) and the AB1048 (2002.2) observation.
The flux drops from the high state to 
the low state at 3.6, 6 and 20~cm.
This provides the shortest time-scale on which the radio observations
show variability.

\section{Interpretation}
\label{section interpretation}

\subsection{Single-star scenario}
\label{section single-star scenario}

\begin{figure*}
\centering
  \includegraphics[scale=0.48]{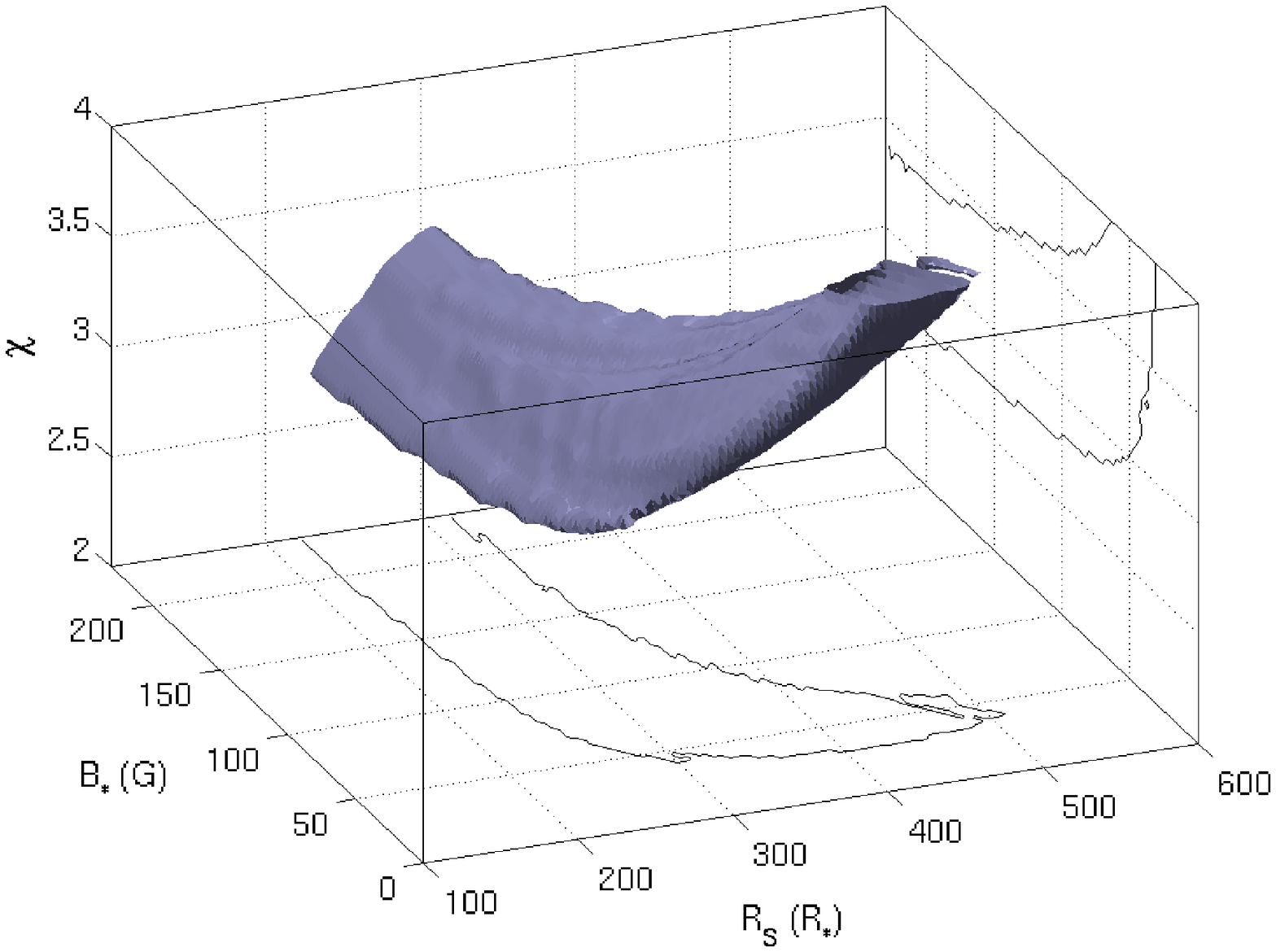}
  \includegraphics[bb=39 189 546 589,scale=0.49]{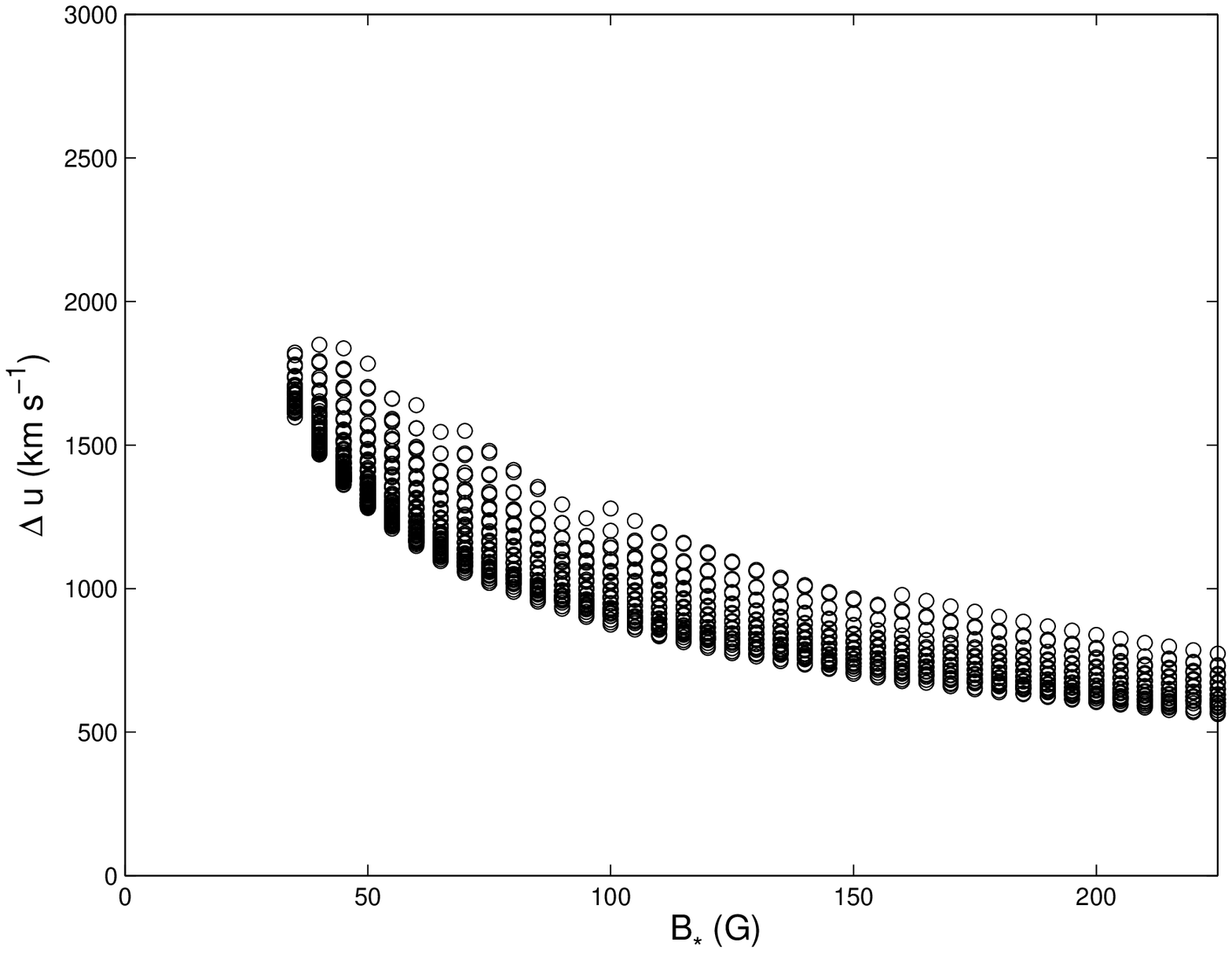}
\caption{(a) {\it Left panel}: The combinations of $\chi$, $R_{\rm S}$ and 
   $B_*$ that fit the AC116 (1984 December 21) observation
   for \object{HD~168112} lie in the grey volume on the
   figure. Projections on two planes are plotted to situate the solutions
   in the parameter space. (b) {\it Right panel}: 
   The values for the surface magnetic
   field $B_*$ and the shock velocity jump $\Delta u$ are shown
   for all possible 
   combinations of the model parameters that explain the
   radio observations of \object{HD~168112}.}
\label{fig:solutionmodel}
\end{figure*}

The non-thermal radiation is due to synchrotron emission by relativistic 
electrons that are Fermi accelerated in shocks. 
In the single-star scenario, the
shocks are wind-embedded (White \cite{White85}) 
as a consequence of the instability of the radiative driving
mechanism. Time-dependent hydrodynamical calculations
confirm that the wind is pervaded by shocks
that can survive to large distances
(e.g. Runacres \& Owocki~\cite{Runacres+Owocki05}, and references therein).
Models for non-thermal emission of O-type stars, based on 
the White~(\cite{White85}) mechanism,
have been developed by Chen \& White~(\cite{Chen+White94}) and
Van Loo et al.~(\cite{VanLoo+al04}, \cite{VanLoo+al05}).
These models are indeed able to explain the observed non-thermal radio
fluxes of \object{Cyg OB2 No. 9} and \object{9 Sgr}, 
though Van Loo et al.~(\cite{VanLoo+al05}) raise some questions
about the high shock velocity jump required
for their fit of the \object{Cyg OB~2 No.~9} data.

In this section, we apply the Van Loo et al.~(\cite{VanLoo+al05})
model to \object{HD~168112}. 
While shocks have a large range of compression ratios 
and velocity jumps, Van Loo et al. showed that the strongest shocks 
dominate the synchrotron emission.
For simplicity, the model assumes a single shock to be
present in the wind. (This assumption is of minor importance
as the results can be re-interpreted in terms of multiple shocks --
see below.) The model is then described by only four
parameters: the position of the shock $R_S$, the compression ratio $\chi$ and 
the velocity jump $\Delta u$ of the shock and the surface magnetic field $B_*$. 

In the model, the momentum distribution of the relativistic electrons
is calculated taking into account the electron acceleration by
the Fermi mechanism (Bell~\cite{Bell78}) and the deceleration
due to inverse-Compton and adiabatic cooling. The synchrotron-emitting
relativistic electrons form a narrow layer downstream
of the shock. The synchrotron flux is then calculated
(including the Razin effect) and corrected for free-free absorption in the wind,
giving the predicted radio flux. We assume that for this star, 
$v_{\rm rot}$ = 250~${\rm km\,s^{-1}}$ (similar to
White~(\cite{White85}) for \object{Cyg OB~2 No.~9})
and that the wind temperature
is 0.3 times the effective temperature (Drew~\cite{Drew89}).
For further details of the model, see
Van Loo et al.~(\cite{VanLoo+al05}).

By calculating a grid of models, we can see which combinations of parameters
explain the observations.
To constrain the fit parameters sufficiently well, it is necessary to have
(nearly) simultaneous radio observations at three or more wavelengths. 
We applied the model to the
AC116 (1984 December 21) observation, which has detections at three 
wavelengths (2, 6 and 20~cm).
Note that this observation is representative of 
the high flux state of \object{HD~168112}.
We calculated the fluxes at the three wavelengths for a grid of $R_S$,
$\chi$, $\Delta u$ and $B_*$. 
The star and wind parameters needed in the model are given in
Table~\ref{table stellar parameters}.
All combinations of the model parameters that fit
the observations within the error bars are shown in 
Fig.~\ref{fig:solutionmodel}.

Figure~\ref{fig:solutionmodel}a shows that only moderately strong to
strong shocks 
($\chi=2.7-4$), located between $220~R_*$ and $500~R_*$ can explain 
the observations. The surface magnetic field is less 
well constrained: we find 
results for $B_*$ from 30~G upward, to values which are above the generally
accepted detection limit for O stars (of order 100~G,
Mathys~\cite{M99}). The most important result, however, is that 
the shock velocity jump is $600 - 1800~{\rm km\,s^{-1}}$ 
(Fig.~\ref{fig:solutionmodel}b).
Such values are considerably higher than those found from time-dependent
hydrodynamical calculations: at $200-500$~$R_*$, $\Delta u$ is typically
only $50~{\rm km\,s^{-1}}$ (Runacres \& Owocki~\cite{Runacres+Owocki05}).

One can lower the required $\Delta u$ by re-interpreting the results of this
single shock model in terms of multiple shocks.
Van~Loo et al. (\cite{VanLoo+al05}) showed that multiple shocks also fit 
the observations provided that the sum of their $\Delta u^3$ equals the single 
shock $\Delta u^3$ (all other parameters being the same). 
To attain the reasonable value of
$\Delta u= 50~{\rm km\,s^{-1}}$ would require more than 1\,700 shocks.
We cannot put these shocks
at arbitrary positions in the 
wind (Van Loo~\cite{VanLoo04}). In the single-shock model,
the resulting 20~cm flux is determined by both the intrinsic synchrotron
radiation and the free-free absorption. If we put the
shocks in a region well beyond the location we found for the
single shock, the free-free absorption will be less effective
and the 20~cm flux will become higher than observed.
We therefore have to put the 1\,700 shocks 
in a region near the location we found for the 
single shock. 
From hydrodynamical calculations, it is obvious that we cannot put that
many shocks in so small a region. 

Another observation suitable for the 
Van Loo et al.~(\cite{VanLoo+al05}) model is the
C978 one, which provides simultaneous observations at 
3.6, 6, 13 and 20~cm. 
The observed fluxes of 3.0, 5.3, 8.2 and 11 mJy
nearly exactly follow a power law. 
(By fitting such a law through the
four data points, we find a spectral index
of $-0.7 \pm 0.1$).
A consequence of this is that there is no
significant upper bound on the shock position $R_S$.
(Only when the larger-wavelength fluxes fall below the power-law
extrapolation of the short-wavelength fluxes can a constraint
on $R_S$ be derived - see Van Loo~(\cite{VanLoo04}).)
The fits show that a large range of shock positions and 
compression ratios fit the observations.
As for the AC116 data, we find that the velocity jump
is too high ($\Delta u \ge 500~{\rm km\,s^{-1}}$) to be compatible
with hydrodynamical models.

We also fitted the observations of 
the AB1065 dataset, which has detections at 3.6 and 6~cm, and a significant
upper limit at 20~cm. Results from
the fits show that the position of the shock is between 50 and 
200~$R_*$. 
The compression ratio is not as well constrained (a consequence
of having only an upper limit at 20~cm): we find $\chi \ge 2$.
Solutions are found for all magnetic field strengths $B_* \ge 10$~G.
Most important, however, is that the $\Delta u$ values are
about the same as for the AC116 dataset.

The observation AB1048 is representative of the low state.
However, no significant constraints can derived from this observation.
The single shock velocity jump should be at least 190~${\rm km\,s^{-1}}$.
This value is lower than for the others because the general flux
level for this observation is low.

The location of the AB1065 shock is more inward than the AC116 shock.
This suggests a direct explanation for the difference in flux level
between the two observations. The AB1065 shock is closer to the star,
so the intrinsic synchrotron flux has to pass through more free-free
absorbing material, resulting in a lower radio flux.
If this is the correct explanation for the variability in the observed
fluxes, we should expect the variability to happen on the hydrodynamical
time-scale, which is of the order of
4 days (the time needed for a shock to travel 
100~$R_*$).
However, the AB671 dataset (Sect.~\ref{section short-term variability}) 
does not
show any variability. To explain why the flux remains constant for nearly
one month, we would have to invoke the fortuitous circumstance of a train
of shocks, all of about equal strength, occurring during one month. 
Ascribing the variability in the
whole dataset to differences in shock positions therefore does not
appear convincing.

As an alternative cause of variability, we could consider variations
in the magnetic field. Very little is known about magnetic
fields in O stars, so it is difficult to make 
predictions on what time-scale
the variability would occur. One possibility is the rotational
time-scale (which is about 1--2 weeks for these stars), but this
is again contradicted by the AB671 observations
(Sect.~\ref{section short-term variability}).

In summary, the single-star scenario can explain the spectral shape,
but has considerable difficulties explaining the flux levels, as a high 
velocity jump or a very large number of shocks would be required. We cannot
attribute the variability to changes in the shock structure. 
In the single-star scenario, we therefore
have to surmise that the variability is caused by an inherent variability
in the magnetic field. The period of the variability is discussed in 
Sect.~\ref{section binary scenario}.

\subsection{Binary scenario}
\label{section binary scenario}

In this section we explore the constraints the data provide assuming
that \object{HD~168112} is a colliding-wind binary.
The best studied example of such a colliding-wind binary is \object{WR~140}
(White \& Becker~\cite{White+Becker95}) and we will use it as
a guide in interpreting the present observations. The 2~cm light
curve of \object{WR~140} shows, during one period, a rising branch followed by
a descending branch. At larger wavelengths, the same feature is seen,
but the peak formed by the two branches becomes sharper. The 
light curve shows good repeatability from one period to another.

The intrinsic synchrotron
radiation will depend on the separation of the two components.
In an eccentric binary more synchrotron emission
will be generated at periastron
than at apastron. However, 
free-free absorption will remove some, or all, of the flux, depending
on the position in the wind (i.e. on the optical depth along the ray
to the observer). The {\em observed} radio lightcurve will therefore
have a maximum at a phase different from periastron. Exactly when the 
maximum occurs depends on the geometry of the orbit.

For close binaries, the colliding-wind region will be completely inside the
free-free radius, and we therefore expect to see no variation on the
time-scale of the orbital period. This is indeed the case for 
\object{Cyg OB~2 No.~5} (Miralles et al.~\cite{Miralles+al94}).
For \object{HD~168112}, however, we may assume that it is not a close binary,
otherwise De Becker et al.~(\cite{De Becker+al04}) would have found
radial velocity variations in their optical spectra, even for an
unfavourable orbital inclination (see also below).  

In the binary hypothesis, we therefore attribute the large radio flux 
changes in Fig.~\ref{figure fluxes} to the orbital motion.
The peaks in flux occur at 1985.0, 1993.1 and 2001.8, which are
$\sim 8.5$~yr apart. The true period is much shorter than this,
as is shown by the rising and descending branches of the lightcurve.
In 1984--5 and 1992--3 we have a rising branch (seen in 2, 3.6 and 6~cm)
lasting $\sim 1$~yr. A descending branch (seen in 3.6, 6 and 20~cm)
is seen around 2002 and lasts $\sim 0.45$~yr. We therefore expect the true
period to be between 1 and 2 years.

\begin{figure}
\resizebox{\hsize}{!}{\includegraphics[bb=28 50 268 311]{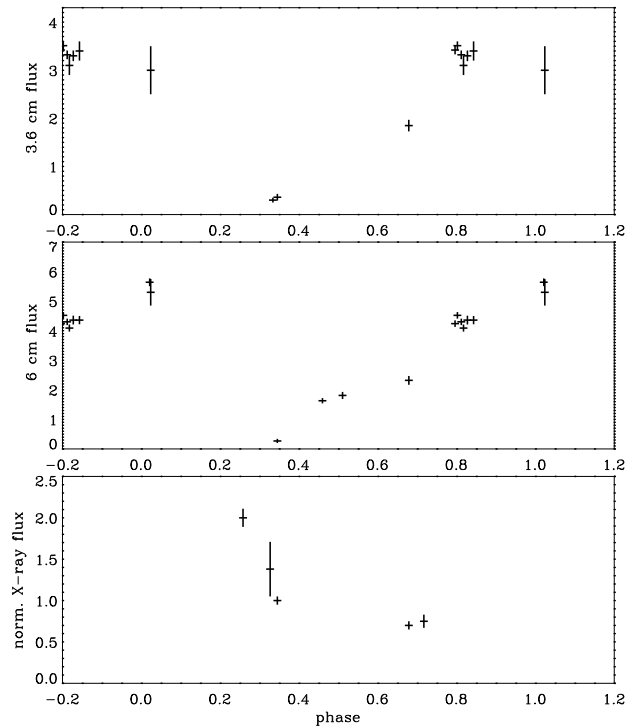}}
\caption{The 3.6 and 6~cm radio fluxes, folded with a 1.4 yr period. 
The normalised X-ray fluxes from De Becker et al.~(\cite{De Becker+al04})
are also plotted (bottom panel). 
Phase 0.0 was arbitrarily set at JD = 2\,444\,000.0.}
\label{figure phases}
\end{figure}

To better determine the period we tried various periods in the range 1--2 yr
and plotted the 3.6 and 6~cm fluxes as a function of phase. (The 2 and
20~cm fluxes do not provide additional information as they are strongly
correlated with the 3.6 and 6~cm fluxes.) We do not have a mathematical
description for the expected behaviour of the fluxes 
as a function of time, so the 
quality of the phase locking was judged by eye. We checked that the
lightcurve shows only one rising and one descending branch over the
whole period and that fluxes at nearby phases have similar values.
Trying a large number of periods we find that $P$ = 1.07, 1.12, 1.22, 
1.4 and 1.66~yr
give good phase locking. The result
for $P=1.4$~yr is shown in Fig.~\ref{figure phases}.

Even the longest period proposed would result in an
orbit of the companion that would
remain within the free-free radius, 
at least if we assume the orbit to be circular.
The orbit with a 1.66~yr period would have a 
semi-major axis of $\sim 90$~R$_*$ (based on the mass listed in
Table~\ref{table stellar parameters} for the primary
and an assumed mass of
30 M$_{\sun}$ for the secondary).
At a radius of $90$~R$_*$, the radial optical depth for free-free absorption
at 20~cm is $\sim$~2.5 (which can be derived from 
Wright \& Barlow~(\cite{Wright+Barlow75}), using
the data from Table~\ref{table stellar parameters}).
This would absorb a large part of the intrinsic radio flux.
Shorter periods would of course suffer even more free-free absorption.
We therefore need to assume an eccentric orbit, so that the colliding-wind
region becomes visible near apastron.
Note that De Becker et al.~(\cite{De Becker+al04})
showed that the X-ray variability of \object{HD~168112} can be explained by
a binary provided the eccentricity is $\ge 0.5$.
An eccentric binary with a 1--2~yr period is also compatible with
the AB671 fluxes being constant
over a one month time-scale. 

X-ray data can provide some further constraints on the period.
Similar to the radio,
more X-ray flux is generated at periastron than at apastron.
For such a wide binary seen under low inclination,
the absorption in X-rays is, however, less than that at
radio wavelengths. 
As the separation between the two components becomes smaller,
the observed X-ray flux therefore increases, while the radio
flux decreases (due to the increasing free-free absorption).
For an eccentric binary, the passage through periastron will be
fast, so we expect a sharp peak in the X-ray emission around that
time.

For \object{HD~168112}, De Becker et al.~(\cite{De Becker+al04}) present two XMM-Newton
observations taken quasi-simultaneously with VLA
radio data (programmes AB1048 and AB1065). 
They also reanalysed three older X-ray observations.
If we fold all the X-ray fluxes with the radio periods found above, we find
that only the 1.4~yr period shows a reasonable behaviour for the
X-ray fluxes. Figure~\ref{figure phases} shows the X-ray data
folded with the 1.4~yr period. The steeply descending
fluxes around phase 0.2--0.4 could be due to the fast change in
separation as the secondary moves away from periastron in the
eccentric orbit. This would suggest that periastron occurs at
phase $\sim$~0.2.

Clumping of the stellar wind material
will not allow a shorter period, or a less eccentric orbit.
This is because clumping has no effect on the optical depth, provided 
that we require the same (thermal) flux to be 
emitted. This can be seen
from Abbott et al.~(\cite{Abbott+al81}). Assuming that the density ratio of 
the low-density versus high-density material is zero, their Eq.~(10)
shows that the flux is proportional to $(\dot{M}^2/f)^{2/3}$, where $f$
is the volume filling factor of the high-density material. The optical
depth (their Eq.~(9)) is proportional to $\dot{M}^2/f$. If we compare a
smooth wind with a clumped wind and require both to have the same
flux, they need to have the same $\dot{M}^2/f$ value, which also means
that the optical depth is the same (for details, see
Van Loo~\cite{VanLoo04}). 
Clumping can also lead to the material becoming 
porous to radiation (Owocki et al.~\cite{Owocki+al04}).
If we would include the effect of porosity as well,
the 20~cm optical depth at $90$~R$_*$ would be smaller, allowing
a shorter period, or a less eccentric orbit for the binary.

One can speculate on the nature of the companion. If it is a normal
star of the same luminosity class as the O5.5 III(f$^+$) primary, 
the spectral type would have to be later than O5.5. 
It would probably not be as late as O9 since otherwise we should see
low excitation lines in the spectrum that would be untypical of the primary.
If the difference
in magnitude becomes larger than 2.5, then the companion could be
of any type 
(a B-star, a pre-main-sequence star of a few solar masses, ...).
A compact companion would produce much brighter X-ray emission than 
what is observed
and the spectrum
would follow a blackbody rather than an optically thin spectrum.
Bieging et al.~(\cite{Bieging+al89}) already dismissed the compact companion
hypothesis as improbable on evolutionary grounds.

Is the presence of a companion compatible with the fact that 
such a companion
was not detected in the available optical spectroscopy? 
We checked this by exploring the parameter space
of possible periods $P$, mass ratios $q=M_2/M_1$, eccentricities $e$
and inclination angles $i$, to see which combinations give
velocity changes that would remain undetected.
We consider a grid of $q$ = 0.1, 0.4, 0.7, 1.0, $e$ = 0.0, 0.5 and
$i$ = 15{\degr}, 30{\degr}, 60{\degr}, 90{\degr}. 
We also took into account the time-scales that are covered by the available
optical data. Adopting an upper limit of 10 km~$\rm{s}^{\rm -1}$ 
(2~sigma) on the semi-amplitude ($K$) of the velocity
variations, we find that for $q \ge 0.4$, 
the inclination should be $i<15\degr$ for a period $P<2$~yr.
For $q=0.1$, higher inclination angles are possible and even
$i=90\degr$ cannot be excluded. 
Of course, there is always the possibility that the optical spectra might have 
been taken at the wrong phase, so that we sample only a small part of
the total velocity variation.

\begin{figure}
\resizebox{\hsize}{!}{\includegraphics[bb=18 144 592 718]{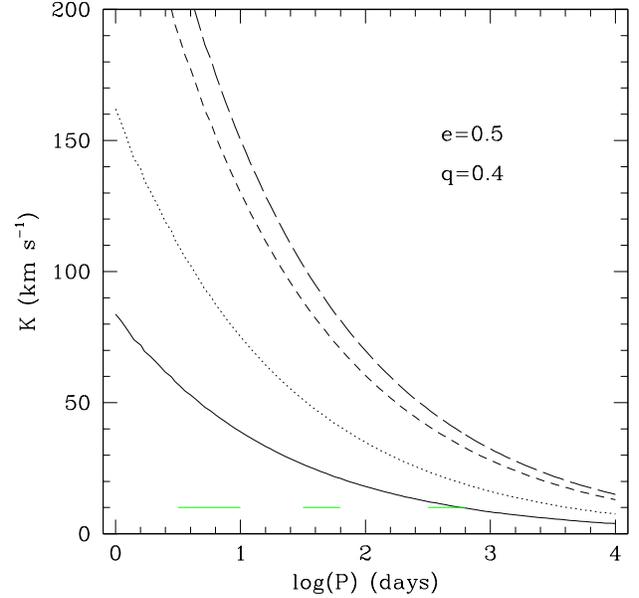}}
\caption{The semi-amplitude $K$ of the velocity
variations as a function of period $P$, for eccentricity $e=0.5$, 
mass ratio $q=0.4$ and inclination $i=15\degr$ (solid line), 
30{\degr} (dotted), 60{\degr} (short-dashed)
and 90{\degr} (long-dashed). The three horizontal lines at the 
10~km~s$^{\rm -1}$ level show the time ranges that have been
covered by the available optical spectroscopy data
(De Becker et al.~\cite{De Becker+al04}).}
\end{figure}

Thus, the observations are in agreement with \object{HD~168112}
being a (low inclination)
binary, with a period of less than 2~yr. It seems that
the most likely period is $\sim$~1.4~yr. The binary scenario
explains the observed variability very well, as shown in
Fig.~\ref{figure phases}. While we did not
model the spectral shape or flux levels, we can derive some
information from the single-star synchrotron model
(Sect.~\ref{section single-star scenario}). Even though that model
is not appropriate for a binary, the location
and the velocity jump of the single shock in that model
are in qualitative agreement
with what can be expected for a colliding-wind region
(e.g. \object{WR~147}, Dougherty et al. \cite{DP03}).

\section{Conclusions}
\label{section conclusions}

In this paper we studied the available VLA and ATCA
radio data on \object{HD~168112}.
The archive data confirm the non-thermal nature and high variability
of the radio emission. Variability is present at a time-scale of
$0.45$~yr or more, and the fluxes can differ by a factor of $\sim 20$.
No variability was found on time-scales less than $0.1$~yr.

We interpreted the observations in terms of both a single-star and
binary scenario. Assuming \object{HD~168112} to be single, we applied
the Van Loo et al.~(\cite{VanLoo+al05}) model to the observations. 
The fact that the strong shocks (which dominate the synchrotron 
emission) are at different positions for different times of observation
was suggested as an explanation for the variability. It fails, however,
to explain why the flux remained constant for about one month.
An even more important
problem with the single-star scenario is that the
shocks need to have velocity jumps very much higher than those
predicted by hydrodynamical models. As an alternative, when we forced
the shocks to have an acceptable velocity jump, we found that too many shocks
are required to explain the flux levels. The failure of the single-star
scenario to give results compatible with hydrodynamical simulations
is a major conclusion of this work.

The binary hypothesis is very successful in explaining all the
observational material. The non-thermal emission is due to 
the colliding winds. The variability can be explained by the
colliding-wind region in an eccentric binary
moving in and out of the region where the
free-free absorption is important. The high value found for the
shock velocity jump is to be expected for a colliding-wind region.
Also, the X-ray data are in agreement
with this scenario. The non-detection by optical spectroscopy
suggests an orbital inclination $i < 15\degr$.
While the amount of radio data available is not large, they clearly
show that the period should be between one and two years. In combination
with the X-ray observations, the present data suggest that the binary
period is most likely $\sim 1.4$~yr. The constraint on the period
is another major
conclusion of this work, and it will allow
a much better planning of future observing campaigns. While the
current evidence favours the binary scenario, confirmation of this
and the determination of the period still need to
be sought from future observations.

\begin{acknowledgements}
We thank Joan Vandekerckhove for his help with the reduction of the
VLA data and Griet van de Steene for discussion on the reduction
of the ATCA data.
We also thank the original observers of the archive data we used.
This research has made use of the SIMBAD database, operated at CDS,
Strasbourg, France and NASA's Astrophysics Data System Abstract Service.
This research has also made use of the NASA/IPAC Extragalactic Database 
(NED) which is operated by the Jet Propulsion Laboratory, California 
Institute of Technology, under contract with the National Aeronautics and 
Space Administration. 
SVL gratefully acknowledges a doctoral research grant by the
Belgian Federal Science Policy Office (Belspo).
Part of this research was carried out in the framework of the
project IUAP P5/36 financed by Belspo.
\end{acknowledgements}

\appendix

\section{Data reduction}
\label{section data reduction}

\subsection{VLA data}
\label{section VLA data}

\begin{table*}
\caption{Reduction of the data at 2, 3.6 and 6~cm. 
Programme C978 is an ATCA observation, all others are 
VLA observations.
Column (1) gives 
the programme name, (2) the date of the observation,
(3) the phase calibrator (J2000 coordinates),
(4) the distance of \object{HD~168112} to the phase calibrator (degrees),
(5) the integration time (in minutes) on the source,
(6) the number of antennas that gave a usable signal,
(7) the configuration the VLA or ATCA was in at the time of the observation,
(8) the major and minor axis (both in arcsec) of the beam,
(9) the position angle (degrees) of the beam,
(10) the RMS in the centre of the image and
(11) the measured flux (in mJy).}
\label{table UCX data}
\begin{center}
\begin{tabular}[]{llllrrrlr@{.}l@{$\times$}r@{.}lrlcl}
\hline
\hline
& \multicolumn{1}{c}{(1)} & \multicolumn{1}{c}{(2)} & \multicolumn{1}{c}{(3)} &
\multicolumn{1}{c}{(4)} & \multicolumn{1}{c}{(5)} & \multicolumn{1}{c}{(6)} &
\multicolumn{1}{c}{(7)} & \multicolumn{4}{c}{(8)} & \multicolumn{1}{c}{(9)} &
\multicolumn{1}{c}{(10)} & \multicolumn{1}{c}{(11)} \\
& \multicolumn{1}{c}{progr.} & \multicolumn{1}{c}{date} &
\multicolumn{2}{c}{phase calibrator} &
\multicolumn{1}{c}{intgr.} &
\multicolumn{1}{c}{no.} &
\multicolumn{1}{c}{config} &
\multicolumn{4}{c}{beamsize} &
\multicolumn{1}{c}{PA} &
\multicolumn{1}{c}{RMS} &
\multicolumn{1}{c}{flux} \\
&  &  &
\multicolumn{1}{c}{name} & 
\multicolumn{1}{c}{dist.} &
\multicolumn{1}{c}{time} & 
\multicolumn{1}{c}{ants.} & &
\multicolumn{4}{c}{(arcsec$^2$)} &
\multicolumn{1}{c}{(deg)} &
\multicolumn{1}{c}{(mJy)} & 
\multicolumn{1}{c}{(mJy)} & \\
\hline
\multicolumn{3}{l}{\bf 2 cm} \\
& AA29   & 1984-04-04 & \object{1733-130} & 11.2 & 22 & 24 & C   &   2&3  &   1&4  & $ 13.9$ & 0.20   & 1.2   $\pm$ 0.3  & \\
& AC116  & 1984-12-21 & \object{1733-130} & 11.2 & 45 & 27 & A   &   0&16 &   0&12 & $  2.2$ & 0.13   & 2.4   $\pm$ 0.2  & \\
\multicolumn{3}{l}{\bf 3.6 cm} \\
& BP1    & 1992-05-30 & \object{1733-130} & 11.2 & 43 & 24 & C   &   9&2  &   4&8  & $-88.8$ & 0.05   & 0.30  $\pm$ 0.05 & \\
& AB671  & 1993-01-21 & \object{1811-209} &  9.0 & 19 & 22 & A   &   0&33 &   0&21 & $ 25.8$ & 0.05   & 3.42  $\pm$ 0.09 & \\
& AB671  & 1993-01-24 & \object{1811-209} &  9.0 & 19 & 27 & A   &   0&41 &   0&23 & $ 38.2$ & 0.05   & 3.51  $\pm$ 0.09 & \\
& AB671  & 1993-01-29 & \object{1811-209} &  9.0 & 19 & 26 & BnA &   0&91 &   0&50 & $ 51.9$ & 0.05   & 3.32  $\pm$ 0.09 & \\
& AB671  & 1993-02-01 & \object{1811-209} &  9.0 &  6 & 27 & BnA &   0&82 &   0&40 & $ 84.4$ & 0.09   & 3.1   $\pm$ 0.2  & \\
& AB671  & 1993-02-06 & \object{1811-209} &  9.0 & 19 & 27 & BnA &   0&83 &   0&42 & $ 68.3$ & 0.04   & 3.30  $\pm$ 0.11 & \\
& AB671  & 1993-02-14 & \object{1811-209} &  9.0 &  9 & 27 & BnA &   0&95 &   0&49 & $ 56.3$ & 0.09   & 3.4   $\pm$ 0.2  & \\
& AB671  & combined   &                   &      &    &    &     &   0&60 &   0&29 & $ 64.6$ & 0.02   & 3.39  $\pm$ 0.07 & \\
& C978   & 2001-10-11 & \object{1832-105} &  3.7 &161 &  6 &EW352&   4&51 &   0&65 & $ -1.3$ & 0.25   & 3.0   $\pm$ 0.5  & \\
& AB1048 & 2002-03-24 & \object{1832-105} &  3.7 & 11 & 26 & A   &   0&31 &   0&25 & $ 11.3$ & 0.06   & 0.36  $\pm$ 0.07 & \\
& AB1065 & 2002-09-11 & \object{1832-105} &  3.7 &  9 & 26 & CnB &   3&0  &   2&1  & $ 47.4$ & 0.11   & 1.85  $\pm$ 0.12 & \\
\multicolumn{3}{l}{\bf 6 cm} \\
& AA28   & 1984-03-09 & \object{1733-130} & 11.2 & 28 & 26 & CnB &   3&7  &   2&1  & $-87.8$ & 0.07   & 1.63  $\pm$ 0.09 & \\
& AA29   & 1984-04-04 & \object{1733-130} & 11.2 & 14 & 27 & C   &   5&9  &   3&7  & $ 10.6$ & 0.11   & 1.81  $\pm$ 0.12 & \\
& AC116  & 1984-12-21 & \object{1733-130} & 11.2 & 50 & 27 & A   &   0&55 &   0&36 & $-20.1$ & 0.06   & 5.64  $\pm$ 0.13 & \\
& AB671  & 1993-01-21 & \object{1811-209} &  9.0 & 19 & 24 & A   &   0&73 &   0&68 & $-27.2$ & 0.07   & 4.24  $\pm$ 0.11 & \\ 
& AB671  & 1993-01-24 & \object{1811-209} &  9.0 & 19 & 27 & A   &   0&55 &   0&44 & $  9.6$ & 0.06   & 4.52  $\pm$ 0.11 & \\ 
& AB671  & 1993-01-29 & \object{1811-209} &  9.0 & 19 & 26 & BnA &   1&7  &   1&1  & $ 34.7$ & 0.06   & 4.30  $\pm$ 0.11 & \\ 
& AB671  & 1993-02-01 & \object{1811-209} &  9.0 &  6 & 27 & BnA &   1&5  &   0&81 & $ 82.5$ & 0.09   & 4.09  $\pm$ 0.12 & \\ 
& AB671  & 1993-02-06 & \object{1811-209} &  9.0 & 18 & 27 & BnA &   1&5  &   1&0  & $ 64.1$ & 0.06   & 4.36  $\pm$ 0.15 & \\ 
& AB671  & 1993-02-14 & \object{1811-209} &  9.0 & 10 & 27 & BnA &   1&7  &   0&92 & $ 51.3$ & 0.08   & 4.36  $\pm$ 0.14 & \\
& AB671  & combined   &                   &      &    &    &     &   0&91 &   0&47 & $ 59.1$ & 0.03   & 4.34  $\pm$ 0.10 & \\
& C978   & 2001-10-11 & \object{1832-105} &  3.7 &161 &  6 &EW352&   7&41 &   1&09 & $ -1.4$ & 0.19   & 5.3   $\pm$ 0.45 & \\
& AB1048 & 2002-03-24 & \object{1832-105} &  3.7 & 11 & 26 & A   &   0&53 &   0&36 & $ 12.3$ & 0.07   & 0.27  $\pm$ 0.07 & \\
& AB1065 & 2002-09-11 & \object{1832-105} &  3.7 &  8 & 26 & CnB &   4&8  &   2&0  & $ 57.4$ & 0.13   & 2.32  $\pm$ 0.15 & \\
\hline
\end{tabular}
\end{center}
\end{table*}

\begin{table*}
\caption{Reduction of the data at 13, 20 and 90~cm. 
Programmes C599 and C978 are ATCA observations, all others are 
VLA observations.
Columns are as in 
Table~\ref{table UCX data}; note that column (10) gives the RMS noise
for the standard reduction in the centre of the image, 
while upper limits in column (11) are based on the
highest RMS found in the search for systematic effects 
(see Appendix~\ref{section VLA data}) and are corrected for the
primary beam effect and bandwidth smearing, if not in the centre
of the image.
Column (12) refers to the notes
and, for the VLA observations,
lists the width of the sidebands if different from $2 \times 50$ MHz.
}
\label{table LP data}
\begin{center}
\begin{tabular}[]{llllrrrlr@{.}l@{$\times$}r@{.}lrr@{.}lr@{.}ll}
\hline
\hline
& \multicolumn{1}{c}{(1)} & \multicolumn{1}{c}{(2)} & \multicolumn{1}{c}{(3)} &
\multicolumn{1}{c}{(4)} & \multicolumn{1}{c}{(5)} & \multicolumn{1}{c}{(6)} &
\multicolumn{1}{c}{(7)} & \multicolumn{4}{c}{(8)} & \multicolumn{1}{c}{(9)} &
\multicolumn{2}{c}{(10)} & \multicolumn{2}{c}{(11)} & \multicolumn{1}{c}{(12)} \\
& \multicolumn{1}{c}{progr.} & \multicolumn{1}{c}{date} &
\multicolumn{2}{c}{phase calibrator} &
\multicolumn{1}{c}{intgr.} &
\multicolumn{1}{c}{no.} &
\multicolumn{1}{c}{config} &
\multicolumn{4}{c}{beamsize} &
\multicolumn{1}{c}{PA} &
\multicolumn{2}{c}{RMS} &
\multicolumn{2}{c}{flux} & \multicolumn{1}{c}{notes} \\
&  &  &
\multicolumn{1}{c}{name} & 
\multicolumn{1}{c}{dist.} &
\multicolumn{1}{c}{time} & 
\multicolumn{1}{c}{ants.} & &
\multicolumn{4}{c}{(arcsec$^2$)} &
\multicolumn{1}{c}{(deg)} &
\multicolumn{2}{c}{(mJy)} & 
\multicolumn{2}{c}{(mJy)} & \\
\hline
\multicolumn{3}{l}{\bf 13 cm}\\
& C978   & 2001-10-11 & \object{1832-105} &  3.7 &120 &  6 &EW352&  15&   &   1&9  & $ -1.3$ &  0&26 & \multicolumn{2}{c}{8.2 $\pm$ 0.6} & X \\
\multicolumn{3}{l}{\bf 20 cm}\\
& BOWE   & 1982-01-31 & \object{1743-038} & 11.9 & 13 &  8 & A   &   4&2  &   1&2  & $ 75.2$ &  7&1  & $<$ 100&   & B,$1 \times 1.5625$ \\
& BAUD   & 1982-02-28 & \object{1743-038} & 11.9 & 33 & 13 & A   &   1&4  &   1&0  & $  5.5$ & 11&   & $<$  55&   & B,$1 \times 1.5625$ \\
& AC116  & 1984-11-27 & \object{1733-130} & 11.2 & 28 & 25 & A   &   1&7  &   1&2  & $  9.4$ &  3&6  & \multicolumn{2}{c}{11   $\pm$ 7}  & B \\
& AC116  & 1984-12-21 & \object{1733-130} & 11.2 & 10 & 27 & A   &   1&7  &   1&2  & $  3.4$ &  0&14 & \multicolumn{2}{c}{8.9 $\pm$ 0.2} & \\
& AG163  & 1984-12-24 & \object{1743-038} & 11.9 &  1 & 27 & A   &   1&8  &   1&2  & $ -9.8$ &  0&47 & $<$  90&   & B,$2 \times 25$ \\
& AC116  & 1985-02-16 & \object{1733-130} & 11.2 & 15 & 25 & A   &   1&6  &   1&2  & $ -2.4$ &  0&12 & $<$  12&   & B \\
& AB531  & 1989-05-01 & \object{1834-126} &  3.9 &  2 & 26 & B   &   6&3  &   4&1  & $ 17.9$ &  1&0  & $<$  26&   & B,$1 \times 50$ \\
& VP51   & 1989-06-19 & \object{1733-130} & 11.2 & 29 & 25 & C   &  15&   &  10&   & $-14.4$ &  0&31 & $<$   1&8  & X \\
& AG290  & 1989-07-20 & \object{1833-210} &  9.6 &  0 & 27 & C   &  22&   &  11&   & $-21.1$ &  1&4  & $<$  60&   & B \\
& VP91   & 1989-11-10 & \object{1733-130} & 11.2 & 18 & 27 & D   &  43&   &  27&   & $-12.2$ &  1&7  & $<$   6&   & X \\
& BP1    & 1992-05-30 & \object{1911-201} & 14.9 & 40 & 25 & C   &  20&   &  15&   & $-70.5$ &  0&80 & $<$   3&8  & \\
& TSTOB  & 1993-07-05 & \object{1822-096} &  2.6 & 14 & 27 & C   &  67&   &  12&   & $ 50.2$ &  0&84 & $<$   3&7  & \\
& AS534  & 1994-05-26 & \object{1822-096} &  2.6 &  5 & 24 & BnA &   4&1  &   2&3  & $-79.5$ &  1&2  & $<$   8&   & B,X,$1 \times 3.125$ \\
& AS534  & 1994-06-01 & \object{1822-096} &  2.6 &  4 & 25 & BnA &   4&9  &   2&1  & $ 56.8$ &  1&6  & $<$  90&   & B,X,$1 \times 3.125$ \\
& AC308  & 1996-06-09 & \object{1833-210} &  9.6 &  1 & 27 & DnC &  38&   &  21&   & $ 81.1$ &  2&4  & $<$  19&   & B \\
& AC308  & 1996-06-16 & \object{1833-210} &  9.6 &  1 & 27 & DnC &  40&   &  20&   & $ 71.7$ &  1&6  & $<$  18&   & B \\
& C599   & 1997-02-25 & \object{1934-638} & 54.0 & 89 &  6 & 6A  &  32&   &   4&4  & $  3.2$ &  0&52 & $<$  23&   & B \\
& AC308  & 1997-10-16 & \object{1833-210} &  9.6 &  1 & 27 & DnC &  38&   &  20&   & $ 84.0$ &  2&2  & $<$  16&   & B \\
& C978   & 2001-10-11 & \object{1832-105} &  3.7 &120 &  6 &EW352&  27&   &   3&4  & $  1.2$ &  0&42 & \multicolumn{2}{c}{11 $\pm$ 2} & X \\
& AB1048 & 2002-03-24 & \object{1834-126} &  3.9 &  5 & 26 & A   &   2&0  &   1&3  & $ 19.0$ &  0&17 & $<$   0&7  &  \\
& AB1065 & 2002-09-11 & \object{1834-126} &  3.9 &  5 & 26 & CnB &   7&1  &   3&3  & $ 40.0$ &  0&68 & $<$   2&8  & I,$\lambda = 18$ cm \\
& AB1065 & 2002-09-11 & \object{1834-126} &  3.9 &  5 & 21 & CnB &   8&1  &   3&8  & $ 38.9$ &  0&37 & $<$   1&5  & $\lambda = 20$ cm\\
\multicolumn{3}{l}{\bf 90 cm} \\
& AH299  & 1988-06-21 & \object{1829+487} & 60.9 & 47 & 21 & DnC & 125&   &  72&   & $-81.0$ & 57&   & $<$ 340&   & B,$2 \times 3.125$ \\
& AH299  & 1989-05-28 & \object{1829+487} & 60.9 & 57 & 25 & CnB &  50&   &  37&   & $-65.7$ & 13&   & $<$  90&   & B,$2 \times 3.125$ \\
\hline
\end{tabular}
\end{center}
Notes:\\
\begin{tabular}{cl}
B & Observation is not centered on \object{HD~168112}. The flux (or upper limit) 
    has been corrected for that. \\
  & No correction for beamwidth smearing
    needs to be applied to C599 (see Sect.~\ref{section ATCA data}). \\
X & The phase calibrator is of low quality (Perley \& Taylor~\cite{Perley+Taylor03}) for VP51, VP91, AS534 and C978. \\
I & Interference is high in the AB1065 18~cm image. \\
\end{tabular}
\end{table*}

The data reduction was done using the Astronomical Image Processing
System (AIPS), developed by the NRAO.
Technical details of the reduction
are listed in Table~\ref{table UCX data} for the 2, 3.6 and 6~cm observations
and Table~\ref{table LP data} for the 20 and 90~cm observations. 
Except where otherwise indicated, all observations were made in 2 sidebands
(denoted IF1 and IF2), each of which has a bandwidth of 50 MHz.
For the flux calibration, we followed the procedure outlined by
Perley \& Taylor~(\cite{Perley+Taylor03}). The fluxes
assigned to the flux calibrators are given in Table~\ref{table calibrators}.
We then determined the instrumental gains from the
phase calibrator observations, interpolated them 
and applied them to the observation of the source. 
Where needed, bad data points were flagged.

\begin{figure*}
\resizebox{\hsize}{!}{\includegraphics[bb=54 360 550 508]{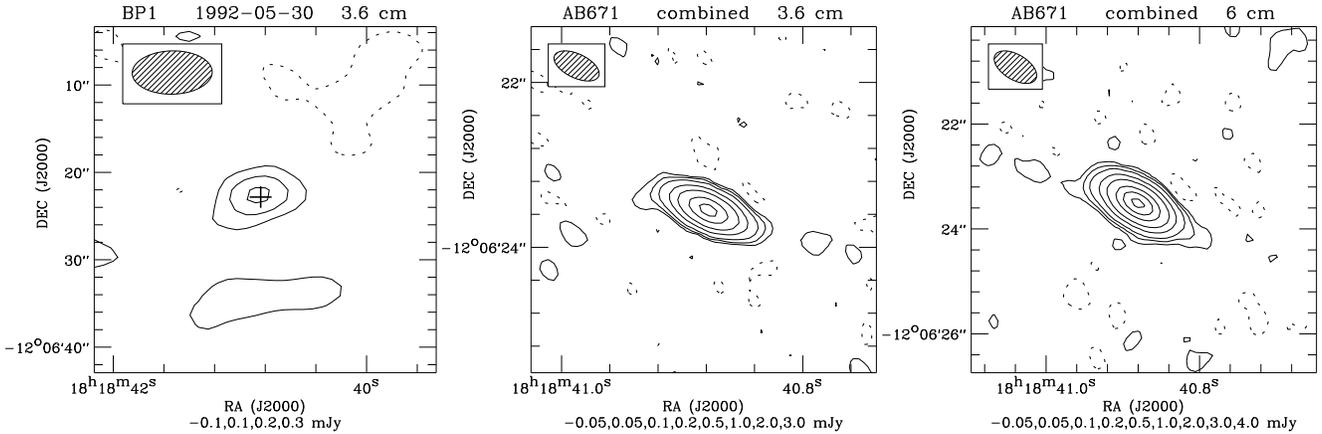}}
\caption{Some examples of maps of VLA archive data 
at 3.6 and 6~cm.
The values of the contour levels
are listed at the bottom of each panel. The negative
contour is given by the dashed line. The first positive contour is at
about twice the RMS noise in the map. 
The beam is shown in the upper left corner.
The cross indicates the optical
ICRS 2000.0 position (from SIMBAD).
Because self-calibration loses absolute positional information,
this is not shown in the middle and right panels.
Note the different spatial scales and different contour levels
of the maps.}
\label{figure maps}
\end{figure*}

The calibrated visibilities
were assigned weights using robust uniform weighting (Briggs~\cite{Briggs95})
and were then inverted by a Fourier transform to produce an image.
For all reductions,
the image covers most or all of the full primary beam.
The image was then deconvolved using the {\sc clean} algorithm
to remove the effect of the beam (point spread function). 
We stopped cleaning
when the algorithm started finding about the same number of negative as
positive components.
On the observations made with low spatial resolution
(VLA configurations C and D), some
Galactic background structure can be seen. We excluded the minimum amount
of observational data taken on the shortest baselines to ensure the
removal of this background.
In Fig.~\ref{figure maps} we show some typical examples of the resulting
maps. 

\begin{table}
\caption{Fluxes of the flux calibrators used in the reduction of 
the archive data. Programmes C599 and C978 are ATCA observations, 
all others are 
VLA observations.
The VLA flux scale is based on
the 1995.2 coefficients (Perley \& Taylor~\cite{Perley+Taylor03}).
The ATCA flux scale is given by Reynolds~(\cite{Reynolds94}).}
\label{table calibrators}
\begin{tabular}{llr@{.}lr@{.}llllllllllll}
\hline
\hline
& calib. & \multicolumn{2}{c}{freq.} & \multicolumn{2}{c}{flux} & programmes \\
&        & \multicolumn{2}{c}{(GHz)} & \multicolumn{2}{c}{(Jy)} \\
\hline
\multicolumn{3}{l}{\bf 2 cm}\\
& \object{3C286} & 14&915 &  3&460 & AA29,AC116 \\
&       & 14&965 &  3&452 & AA29,AC116 \\
\multicolumn{3}{l}{\bf 3.6 cm}\\
& 3C286 &  8&415 &  5&224 & AB671,BP1 \\
&       &  8&465 &  5&203 & AB671 \\
& \object{3C48}  &  8&435 &  3&164 & AB1048,AB1065 \\
&       &  8&485 &  3&145 & AB1048,AB1065 \\
& \object{1934-638} &8&640&  2&841 & C978 \\
\multicolumn{3}{l}{\bf 6 cm}\\
& 3C286 &  4&835 &  7&510 & AA28,AA29,AC116,AB671 \\
&       &  4&885 &  7&462 & AA28,AA29,AC116,AB671 \\
& 3C48  &  4&835 &  5&459 & AB1048,AB1065 \\
&       &  4&885 &  5&405 & AB1048,AB1065 \\
& 1934-638&4&800 &  5&828 & C978 \\
\multicolumn{3}{l}{\bf 13 cm}\\
& 1934-638&2&496 & 11&14  & C978 \\
\multicolumn{3}{l}{\bf 20 cm}\\
& 3C286 &  1&365 & 15&01 & AC308  \\
&       &  1&435 & 14&65 & AC308  \\
&       &  1&452 & 14&57 & AG163  \\
&       &  1&465 & 14&51 & AC116,AG290,AB531 \\
&       &  1&502 & 14&34 & AG163  \\
&       &  1&515 & 14&28 & AC116,AG290 \\
&       &  1&612 & 13&85 & BAUD,BOWE,AS534   \\
&       &  1&665  & 13&63 & VP91,BP1\\
& 3C48  &  1&385 & 16&20 & AB1048,AB1065,TSTOB \\
&       &  1&465 & 15&49 & AB1048,AB1065 \\
&       &  1&635 & 14&19 & AB1065 \\
&       &  1&665 & 13&98 & AB1065 \\
& \object{1733-130} & 1&665& 5&200   & VP51$^{\mathrm{a}}$ \\
& 1934-638&1&384 & 14&94  & C599,C978 \\
\multicolumn{3}{l}{\bf 90 cm}\\
& 3C286 &  0&328 & 25&90 & AH299 \\
&       &  0&333 & 25&77 & AH299 \\
\hline
\end{tabular}
\begin{list}{}{}
\item[$^{\mathrm{a}}$] for VP51, no flux calibrator was available, 
so we calibrated
the flux on the phase calibrator, to which we assigned the value listed
in Perley \& Taylor~(\cite{Perley+Taylor03}).
\end{list}
\end{table}

We measured the fluxes on the cleaned image using the AIPS task
{\sc jmfit} to fit elliptical Gaussians to the source.
The elliptical
Gaussian was forced to have the major axis, minor axis and position
angle equal to the beam values.
For the 20~cm AC116 (1984-11-27) observation, \object{HD~168112} is not
close to the centre of the map. In that case,
the flux measurement becomes more complicated. 
First, the sensitivity of the image decreases as
one moves away from the centre (``primary beam attenuation").
The images can easily be corrected for this effect
(using the task {\sc pbcor}). A second effect is
beamwidth smearing (Bridle \& Schwab~\cite{Bridle+Schwab99}):
an off-centre source is smeared out in the direction that is radial
toward the centre of the image.
Fortunately, this effect conserves
the total flux. For this observation, we measured the flux by using
an elliptical Gaussian with all parameters left free.
As a check, we used the interactive task {\sc tvstat} to integrate
the flux over the area occupied by the beamwidth-smeared source.
We found that both measurements were in acceptable agreement.

Usually, the measured fluxes are assigned an error bar corresponding
to the root-mean-square (RMS) noise in the map. 
The RMS error bar only covers the random sources of error. 
To estimate some of 
the {\em systematic} errors, we redid the reduction a number of times.
We excluded one antenna from the data, excluded one run  (if there were at
least three runs),
tried natural or uniform weighting instead of robust uniform weighting or
varied the extent of the taper.
The error bar we assign to the measured fluxes covers the range
in fluxes we thus found. In many
cases this is larger than the RMS error bar.
We also measured the flux by determining the maximum intensity of the point
source, instead of fitting an elliptical Gaussian.
In most cases, both measurements agree within the error bars. The
worst non-agreement is the 6~cm AC116 observation, where the maximum
intensity is $-2.5$~$\sigma$ below the integrated value. Lesser effects
are shown by AB1065 -- 3.6~cm ($+1.2$~$\sigma$), AA28 -- 6~cm ($-1.4$~$\sigma$)
and AB671 1993-01-24 -- 6~cm ($-1.5$~$\sigma$).

The final error bar was determined by adding (in a RMS sense)
a 2 \% calibration error
for the 3.6, 6, 20 and 90~cm observations and a 5 \% error for the 
2~cm. These represent the uncertainty in the absolute
calibration (Perley \& Taylor~\cite{Perley+Taylor03}).

For those observations where \object{HD~168112} was not detected, we assigned an 
upper limit of 3 times the RMS noise measured in the map (in a small box 
around the source). To take into account some of the systematic effects,
we also performed the various reductions described above and used 
the highest RMS thus found to define the upper limit.
When \object{HD~168112}
is offset from the centre, the RMS must be measured on the image corrected
for primary beam attenuation.
To correct for the beamwidth smearing,
we had to multiply the RMS
by the number of beams that a detectable
beamwidth-smeared source at this position would cover.
From Bridle \& Schwab~(\cite{Bridle+Schwab99})
we find that the number of beams can be approximated
by $\sqrt{1+\beta^2}$, where:
\begin{equation}
\beta = \frac{\mathrm{bandwidth~(MHz)}}{\mathrm{observing~frequency~(MHz)}} 
    \times
\frac{\mathrm{offset~(arcsec)}}{\mathrm{beamwidth~(arcsec)}}.
\end{equation}

In some cases \object{HD~168112} is within the primary beam
on two images centred on different
targets. In that case we only list the one closest to \object{HD~168112} because
this gives the smallest error bar. We did check that the flux of the
other image is compatible with the flux listed. The following observations
were not listed because they are of such low quality 
that they do not even provide a significant upper limit:
BAUD (1980-07-12; 20~cm), FIX (1982-05-22; 20~cm), 
AC161 (1986-05-11; 90~cm),
AH250 (1986-12-27 and 1987-03-19; 90~cm),
AK162 (1987-03-20; 90~cm),
AH299 (1988-03-24, 1988-05-02 and 1988-09-23; 90~cm),
AT143 (1993-06-11; 20~cm),
AL341 (1994-09-13; 90~cm) and
AG626 (2002-08-14, 2002-11-23, 2002-12-07, 2003-02-23 and 2003-02-24;
90~cm).

A complication arose in the reduction of the TSTOB data. 
Comparing the fluxes of other sources in the field around \object{HD~168112}
across the various programmes showed that the
TSTOB values are too low by a factor of $\sim 2$. Fortunately, the TSTOB
programme was immediately followed by another programme (AW362) that
observed the same flux calibrator at the same wavelength. Using the
flux calibrator from that programme 
in the reduction, the TSTOB fluxes become
high enough to be compatible with the others. The results in
Fig.~\ref{figure fluxes} and Table~\ref{table LP data} are based
on this improved reduction.

A comparison with the literature (Bieging et al.~\cite{Bieging+al89},
Phillips \& Titus~\cite{Phillips+Titus90},
De Becker et al.~\cite{De Becker+al04})
shows acceptable agreement in most cases.
There is a tendency for our error bars (or upper limit
in the case of Phillips \& Titus) to be larger:
this is due to our inclusion of systematic effects. 
Note also that the AB1048 and AB1065 fluxes are somewhat different
from those published in De Becker et al.~(\cite{De Becker+al04}).
This is because, in the present paper, we followed the recommendation by 
Perley \& Taylor~(\cite{Perley+Taylor03}) to assign
a small, but non-zero, weight ({\sc wtuv} = 0.1)
to the flux calibrator visibilities outside the usable range.
There are two exceptions to the acceptable agreement.
For the 2~cm AC116 (1984-12-21) observation, we have $2.4 \pm 0.2$ mJy, 
while Bieging et al. find $1.3 \pm 0.1$ mJy. They note an apparent increase 
of flux at the largest baselines. Our reduction of the data does not show such
a problem, giving a higher confidence in our result.
For the 6~cm AA28 (1984-03-09) observation we found $1.63 \pm 0.09$ mJy,
while Bieging et al. list $1.3 \pm 0.1$ mJy. 
We could not find an explanation for this discrepancy.

\subsection{Self-calibration}
\label{section AB671 self-calibration}

The AB671 dataset has a number of problems.
The phase calibrator used has a low 3.6~cm flux, making it 
unsuitable for phase calibration.
The time between two phase calibrations exceeds 30 min in 5 of
the 6 observations, which is too long to follow the phase changes
due to the atmosphere, especially in these high-resolution observations. 
When we applied our standard reduction, we
found considerable variability in the fluxes. However,
these flux variations are highly correlated with the flux variations of
\object{HD~167971}, 
which was observed as part of the same observing programme.
At least a large part of the variability must therefore be
due to atmospheric effects that were not correctly removed
in the standard reduction.

We therefore decided to apply phase self-calibration 
(see, e.g. Cornwell \& Fomalont~\cite{Cornwell+Fomalont99})
to this set of observations. This technique is suitable
for the ``simple" field here, consisting of two point sources
on an empty background.
It allows us to follow the atmospheric phase changes on a shorter
time-scale than the time between two phase calibrator observations.
From the known sensitivity of the VLA and the estimated flux of \object{HD~168112}, 
we find that a 3 min time-scale is appropriate, provided that we combine 
both sidebands and all Stokes polarizations and decrease the requested 
signal-to-noise ratio of the gain solutions to 2 
(Ulvestad~\cite{Ulvestad02}). Ideally, we would prefer a 10 sec time-scale
(which is the duration of a single integration), so we could follow the
phase changes even better. Unfortunately, insufficient signal would then
be available for the calibration.
We applied 5 iterations of phase-only calibration. No amplitude
self-calibration was done because amplitude errors are relatively 
unimportant on the dynamic ranges of the images we have.
The results presented in Figs.~\ref{figure fluxes}
and \ref{figure AB671 fluxes}
and Table~\ref{table UCX data} are those based on the self-calibration
reduction.

To make sure that the self-calibration technique does not introduce
systematic effects, we also applied it to those cases where 
it is not really needed. These tests show that self-calibration and 
the standard
reduction give the same results within a few percent, provided that 
the self-calibration is carefully applied (i.e. not on a complicated background,
no interference present in the image and 
a time-scale appropriate for the flux).
A side effect of self-calibration is that the absolute positional
information is lost.

\subsection{ATCA data}
\label{section ATCA data}

\begin{figure}
\resizebox{\hsize}{!}{\includegraphics[bb=54 360 309 592]{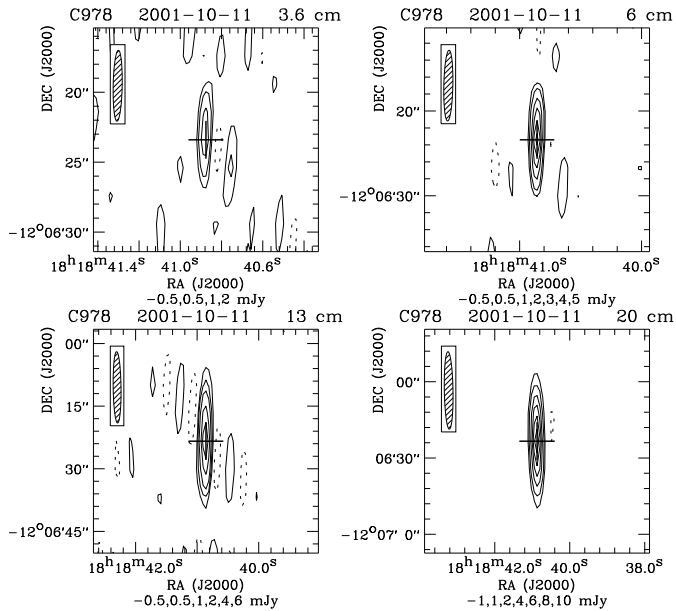}}
\caption{Maps of ATCA archive data where \object{HD~168112} was detected.
The values of the contour levels
are listed at the bottom of each panel. The negative
contour is given by the dashed line. The first positive contour is at
about twice the RMS noise in the map. 
The beam is shown in the upper left corner.
The cross indicates the optical
ICRS 2000.0 position (from SIMBAD).
Note the different spatial scales and different contour levels
of the maps.}
\label{figure maps ATCA}
\end{figure}

The ATCA data were reduced in Miriad following the user guide 
(Sault \& Killeen~\cite{Sault+Killeen99}).
We did not reduce the 13~cm data of the C599 programme because
\object{HD~168112} is well outside the primary beam.
The data were read into Miriad and corrected for self-interference of the 
array and for the phase difference between the X and Y
channels. 
After reading in, the data consist of 13 frequency channels, each
of 8 MHz width. The reduction then proceeds in a way similar to the
VLA reduction (Appendix~\ref{section VLA data}). 
The use of multiple frequency channels requires us to do a bandpass
calibration on the phase calibrator.

Fluxes assigned to the
flux calibrator are listed in Table~\ref{table calibrators}.
The programme C978 uses the 
phase calibrator \object{1832$-$105} (J2000), 
which is not recommended at 20~cm
(according to the VLA Calibrator Manual, 
Perley \& Taylor~\cite{Perley+Taylor03}).
For the 13 and 20~cm observations, 
we therefore made a map around this calibrator. 
To remove structure in the background, we had to flag the shortest baselines.
In our data, the phase calibrator turns out to be a point source (with a flux
of 1.04~Jy at 13~cm and 0.94~Jy at 20~cm). There are no other sources
in the field that have a flux higher than 0.01~Jy. Thus, we consider
\object{1832$-$105} to be an acceptable calibrator for our data.

For the target observation, we also had to exclude data from the 
shortest baselines,
to remove background structure in the resulting image.
In the inversion and cleaning step, we
applied multi-frequency synthesis (MFS), which
compensates for the spectral index of the source across the bandwidth.
An interesting consequence of MFS is that we do not have to correct
for bandwidth smearing for those targets that are offset from the
centre (Cotton~\cite{Cotton99}).
The cleaning was applied in a box around the target.
At larger wavelengths more sources are visible on the image, and more
boxes were therefore used. We stopped cleaning
when the algorithm started finding about the same number of negative as
positive components. Applying this criterion, we arrive close to the
theoretical RMS, which is the noise calculated taking only
the system temperature of the front-end
receiver into account, not the calibration errors, side-lobes or any
other instrumental effects. Only for the 20~cm observation do we
end up with an RMS level of about twice the theoretical RMS. We
attribute this to the more complicated background at this wavelength.
Results are listed in Tables~\ref{table UCX data} and
\ref{table LP data}.
These results include the flux calibration uncertainty of 2 \%
and the estimated effect of systematic errors (derived in a similar way
as for the VLA data). The detections are plotted in
Fig.~\ref{figure maps ATCA}.


\begin{thebibliography}{}
\bibitem[1981]{Abbott+al81} Abbott, D. C., Bieging, J. H., \& Churchwell, E.
               1981, ApJ, 250, 645
\bibitem[1985]{Abbott+al85} Abbott, D. C., Bieging, J. H., \& Churchwell, E.
               1985, in Radio Stars, eds. R. M. Hjellming, \& D. M. Gibson,
               D. Reidel, Dordrecht, 219
\bibitem[1978]{Bell78} Bell, A. R. 1978, \mnras, 182, 147
\bibitem[1989]{Bieging+al89} Bieging, J. H., Abbott, D. C., \& Churchwell, 
               E. B. 1989, ApJ, 340, 518
\bibitem[1999]{Bridle+Schwab99} Bridle, A. H., \& Schwab, F. R. 1999,
               in Synthesis Imaging in Radio Astronomy II, eds. G. B. Taylor,
               C. L. Carilli \& R. A. Perley, ASP Conf. Ser., 180, 371
\bibitem[1995]{Briggs95} Briggs, D. S. 1995, High Fidelity Deconvolution of
               Moderately Resolved Sources, PhD thesis (The New Mexico 
               Institute of Mining and Technology, Socorro, New Mexico) \\
               {\tt http://www.aoc.nrao.edu/dissertations/dbriggs}
\bibitem[2002]{Cappa+al02} Cappa, C. E., Goss, W. M., \& Pineault, S.
               2002, AJ, 123, 3348
\bibitem[1994]{Chen+White94} Chen, W., \& White, R. L. 1994, Ap\&SS, 221, 259
\bibitem[1977]{Conti+Ebbets77} Conti, P. S., \& Ebbets, D. 1977, ApJ, 213, 438
\bibitem[1999]{Cornwell+Fomalont99} Cornwell, T., \& Fomalont, E. B. 1999,
               in Synthesis Imaging in Radio Astronomy II, eds. G.B. Taylor,
               C.L. Carilli \& R.A. Perley, ASP Conf. Ser., 180, 187
\bibitem[1999]{Cotton99} Cotton, W. D. 1999,
               in Synthesis Imaging in Radio Astronomy II, eds. G. B. Taylor,
               C. L. Carilli \& R. A. Perley, ASP Conf. Ser., 180, 357
\bibitem[2004]{De Becker+al04} De Becker, M., Rauw, G., Blomme, R., et al.
               2004, A\&A, 420, 1061
\bibitem[2000]{Dougherty+Williams00} Dougherty, S. M., \& Williams, P. M.
               2000, MNRAS, 319, 1005
\bibitem[2003]{DP03} Dougherty, S. M., Pittard, J. M., Kasian, L., et al. 
               2003, \aap, 409, 217
\bibitem[1989]{Drew89} Drew, J. E. 1989, ApJS, 71, 267
\bibitem[1993]{Eichler+Usov93} Eichler, D., \& Usov, V. 1993,
               ApJ, 402, 271
\bibitem[1973]{Georgelin+al73} Georgelin, Y. M., Georgelin, Y. P., \& Roux, S.
               1973, A\&A, 25, 337
\bibitem[1988]{Leitherer88} Leitherer, C. 1988, ApJ, 326, 356
\bibitem[1999]{M99} Mathys, G. 1999, in Proc. IAU Coll. 169, 
               Variable and Non-spherical Stellar Winds in Luminous Hot 
               Stars, eds. B. Wolf, O. Stahl, \& A. W. Fullerton,
               Lect. Notes Phys., 523, 95
\bibitem[1994]{Miralles+al94} Miralles, M. P., Rodr\'{\i}guez, L. F., 
               Tapia, M., et al. 1994, A\&A, 282, 547
\bibitem[1988]{Owocki+al88} Owocki, S. P., Castor, J. I., \& Rybicki, G. B.
               1988, ApJ, 335, 914
\bibitem[2004]{Owocki+al04} Owocki, S. P., Gayley, K. G., \& Shaviv, N. J.
              2004, ApJ, 616, 525
\bibitem[2003]{Perley+Taylor03} Perley, R. A., \& Taylor, G. B. 2003,
               The VLA Calibrator Manual \\
               {\tt http://www.aoc.nrao.edu/$\sim$gtaylor/calib.html}
\bibitem[1990]{Phillips+Titus90} Phillips, R. B., \& Titus, M. A. 1990,
               ApJ, 359, L15
\bibitem[1994]{Reynolds94} Reynolds, J. 1994, A Revised Flux Scale
              for the AT Compact Array, ATNF Internal Report,
              AT/39.3/040\\
              {\tt http://www.atnf.csiro.au/observers/memos/d96783$\sim$1.pdf}
\bibitem[2005]{Runacres+Owocki05} Runacres, M. C., \& Owocki, S. P. 2005, 
               \aap, 429, 323
\bibitem[1999]{Sault+Killeen99} Sault, B., \& Killeen, N. 1999,
              Miriad Users Guide \\
              {\tt http://www.atnf.csiro.au/computing/software/miriad}
\bibitem[2002]{Ulvestad02} Ulvestad, J. 2002, Eighth Synthesis Imaging 
               Summer School, 18-25 June 2002, Socorro, New Mexico, USA \\
               {\tt http://www.aoc.nrao.edu/events/synthesis/2002}
\bibitem[2005]{VanLoo04} Van Loo, S. 2005, PhD thesis, in preparation
\bibitem[2004]{VanLoo+al04} Van Loo, S., Runacres, M. C., \& 
                Blomme, R. 2004, A\&A, 418, 717
\bibitem[2005]{VanLoo+al05} Van Loo, S., Runacres, M. C., \& 
                Blomme, R. 2005, A\&A, in press
\bibitem[1985]{White85} White, R. L. 1985, \apj, 289, 698
\bibitem[1995]{White+Becker95} White, R. L., \& Becker, R. H. 1995, 
               ApJ, 451, 352
\bibitem[1975]{Wright+Barlow75} Wright, A. E., \& Barlow, M. J. 1975,
               MNRAS, 170, 41
\end{thebibliography}
\end{document}